\documentclass[]{aastex}
\usepackage{emulateapj5}

\usepackage{enumerate}
\usepackage{epsfig,psfig,epsf}

\newcommand{\GG}{GRS~1915+105}
\newcommand{\GRO}{GRO~J$1655$-$40$}

\newcommand{\eg}{{\em e.g.}  }
\newcommand{\etalb}{{\em et al.}}

\newcommand{\ie}{{\em i.e.}  }
\newcommand{\Oint}{$\Omega_{int}$} 
\newcommand{\rint}{$r_{int}$} 
\newcommand{\ltsima}{$\; \buildrel < \over \sim \;$}
\newcommand{\simlt}{\lower.5ex\hbox{\ltsima}}
\newcommand{\gtsima}{$\;\buildrel>\over\sim\;$}
\newcommand{\simgt}{\lower.5ex\hbox{\gtsima}}
\shorttitle{Magnetic floods in \GG}
\shortauthors{Tagger, Varni\`ere, Rodriguez and Pellat}
\begin{document}


\title{Magnetic floods: a scenario for the variability \\ of the microquasar \GG}
\submitted{accepted for publication in ApJ}

\author{M. Tagger}
\affil{Service d'Astrophysique (CNRS FRE 2591), CEA Saclay, France\\
and\\
F\'ed\'eration de Recherche Astroparticule et Cosmologie, Universit\'e Paris VII}
\email{tagger@cea.fr}
\author{P. Varni\`ere}
\affil{Department of Physics and Astronomy, University of Rochester, Rochester, NY 14627}
\author{J. Rodriguez}
\affil{Service d'Astrophysique (CNRS FRE 2591), CEA Saclay, France\\
and\\
Integral Science Data Center, Chemin d'Ecogia, 16, CH-1290 Versoix, Switzerland}
\author{R. Pellat\altaffilmark{1} }
\affil{Centre de Physique Th\'eorique, Ecole Polytechnique, Palaiseau, France\\
and\\
Physics Department, UCLA}
\altaffiltext{1}{Deceased August 4, 2003.}
\begin{abstract}
We present a scenario for the variability of the microquasar \GG. This
starts from  previous works, leading to the tentative 
identification of the Accretion-Ejection Instability as the source of
the low-frequency Quasi-Periodic 
Oscillation of microquasars and other accreting sources. We follow the
physics of this instability: 
its conditions (the magnetic field and geometry adapted to MHD jet
models), its instability criterion, 
and its consequences (cooling down of the disk, heating and excitation
of the corona). Comparing 
them to the observed properties of the source, in particular the
detailed properties of its spectral 
states, we first derive a model for the $\sim$ 30 minutes cycles often
exhibited by \GG.  In our 
model this is a limit cycle determined by the advection of poloidal
magnetic flux to the inner 
region of the disk, and its destruction by reconnection (leading to
relativistic ejections) 
with the magnetic flux trapped in the vicinity of the central
source. We show how it leads to natural explanations for observed behaviors of \GG, including the  three basic states of \citet{Bel00}. We then discuss how 
this could be extrapolated further to understand the longer-term
variability of this and 
other microquasars.

\end{abstract}




\section{Introduction}\label{sec:Intro}
Since they were observed in X-ray binaries, Quasi-Periodic
Oscillations (QPO) have been considered as an important clue to the
physics of the inner region of accretion disks, at a few tens to
hundreds of kilometers from the central source (a white dwarf, a
neutron star or a black hole). Their frequencies, typical of keplerian
rotation in this region, and their RMS amplitude (typically a few \%,
up to a few tens of \%, of the total luminosity), as well as a number of other elements, tend
to indicate that they are associated with high magnitude phenomena
occurring in the disk, and possibly at the base of the jet. However no
model has yet gained general acceptance to explain any of the QPO, or
given indications for a more general explanation of the accretion
process. \\
We have in the last few years presented the Accretion-Ejection
Instability (AEI: \citet {TP99}) and shown that its properties make it
a very promising explanation for the low-frequency (\(\sim\) 1-10 Hz)
QPO (hereafter LFQPO), prominent in black-hole binaries and also
present in certain states of those hosting a neutron star or a white
dwarf.  The goal of this paper is to present a model which was
elaborated gradually, since the discovery of the AEI; earlier versions
of this model were presented in \cite {Tag00} and \citet{Tag02}. It
starts from the tentative identification of the AEI as the source of
the QPO, and extrapolates by seeking in the properties of the
instability an explanation for the observed behavior of the most
spectacular microquasar, \GG. \\
Since its discovery \citep{Cas92, Cas94}, soon followed by the
observation of massive superluminal ejections \citep{Mir94}, this source has
been the object of a large number of observations, from X-rays (mostly
by the Rossy X-ray Timing Explorer, RXTE), to radio and IR. They have
characterized it as the most permanently active black-hole binary,
showing a large variety of temporal and spectral behaviors. This occurs at all time scales: high and low-frequency QPOs, but also repetitive patterns of behaviors during typically minutes or tens of minutes, a very long and steady \lq plateau\rq{} state, etc. \cite{Bel00}  have given a classification of these modes of variability, showing that they fall into twelve distinctive classes which again can be reduced to 
oscillations between three basic states. These behaviors are so well characterized and  so well observed at all wavelengths (including their connection with both a compact jet and relativistic ejections) that they 
are widely expected to contain clues to the accretion and ejection
processes in all types of accreting sources, from young stars to X-ray
binaries and AGN. We will discuss this in more details in sections \ref{sec:GG} and \ref{subsec:longterm}, and clearly we expect that our model will be found relevant to other accreting sources.\\
In particular in this paper we concentrate on the cycles where, with a periodicity of the
order of 30 minutes, \GG{} alternates between a high/soft and a
low/hard state, and produces relativistic ejections. As shown by
\cite{Bel00}, this cycle is the most frequent type of variability
observed in this source besides a steady hard state. Our `Magnetic Flood'
model owes its name to the fact that we are led, in our extrapolation
process, to believe that these cycles, and maybe the longer-term
behaviour of \GG, are controlled by the accumulation and release of
poloidal magnetic flux in the innermost region of the accretion
disk. \\
In this sense our scenario has much in common with the model of
\citet{Liv03}, who propose that accreting sources switch between two
states, controlled by the global poloidal magnetic field. This follows
in particular the work of \citet{Mer03}, who finds that the low-hard
state of \GG{} can be fitted by assuming that a substantial fraction
$f$ of the accretion energy is emitted to the corona and jet rather
than heating the disk. Our work is different in that we start from a
definite physical model of accretion in the disk, due to the AEI, and
build up on that. This leads to a number of differences, \eg as will
be seen in section \ref{subsec:states} in the nature of state {\em A}
as defined by \citet{Bel00}. \\
Beyond the identification of the AEI as the source of the LFQPO, our
model is built by extrapolation, and does not rest on precise
arguments, such as predictions or numerical values.  It only claims to
be a possible explanation for the observed behaviors, obtained by
following the most straightforward line of deductions. On the other
hand, as will be seen in this paper, we find it already useful in
guiding us as we seek new diagnostics of the disk physics. It is
however encouraging that, since it was first proposed, we have found
it compatible with new elements gathered in the observations, and that
we obtain with it possible explanations for puzzling observations, such as the transitions between Belloni's states \citep{Bel00}. \\
The paper is organized as follows: in section \ref{sec:AEI}, so that this paper is reasonably self-contained, we review theoretical and numerical results on the AEI and its main properties. Section \ref{sec:GG} is dedicated to a short review of the
variability of \GG, from the time scale of the QPO to that of the
$\sim$ 30 minutes cycles and to properties observed on much longer
periods. Section \ref{sec:flood} will present our model, and section
\ref{sec:discu} our conclusions.
\section{The Accretion-Ejection Instability}\label{sec:AEI}
\subsection{Basic theory}\label{subsec:theo}
\begin{figure*}
\centering
\epsfig{file=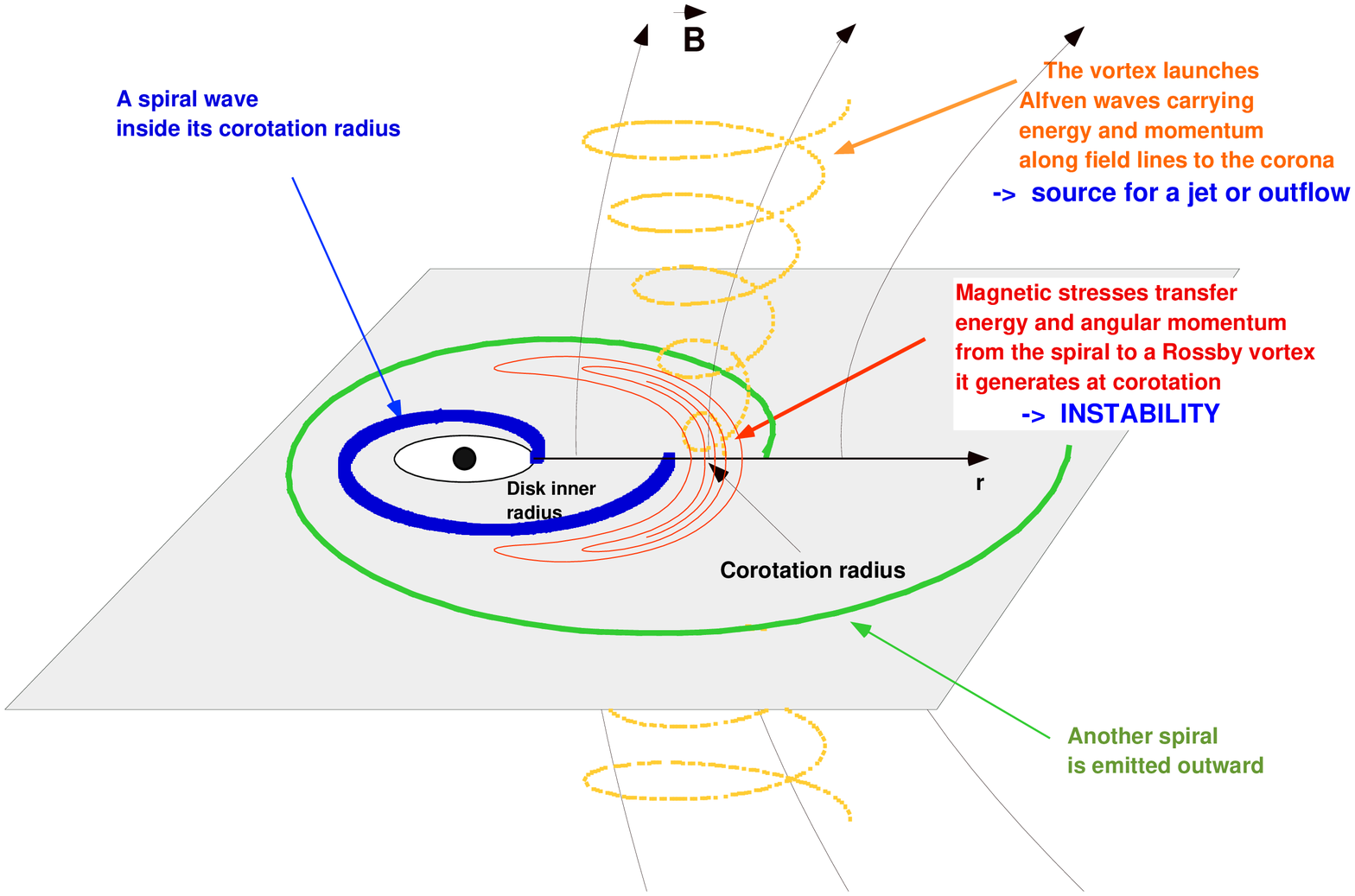,width=\columnwidth}
\caption{A schematic description of the structure of the AEI in the
disk and corona. As other spiral modes (self-gravitating galactic
disks, Papaloizou-Pringle instability) it is formed essentially by
spiral waves traveling back and forth between the inner disk edge (or
the galactic center) and the corotation radius, where the waves rotate
at the same speed as the gas.  The AEI  mode is mainly amplified by coupling to a Rossby vortex it generates at the corotation radius, and transferring to it the energy and angular momentum it extracts from the disk. In a disk threaded by a poloidal magnetic field, the torsion associated with the vortex propagates along the field lines as Alfv\'en waves, carrying to the corona a substantial fraction of the
accretion energy extracted by the spiral from the disk.
}\label{fig:AEI}
\end{figure*}
The Accretion-Ejection Instability \citep{TP99} is a global mode occurring in
magnetized disks threaded by a poloidal (vertical) magnetic field, \ie the geometry used in MHD jet models. It belongs to the same family as the spiral instability of
self-gravitating disks, and the Papaloizou-Pringle instability
(hereafter PPI) of unmagnetized, non-self-gravitating ones. It can
also be viewed as the p-mode of diskoseismology models \citep{Now91},
modified by magnetic stresses; the big difference here is that as in
galaxies (though less violently) this mode is {\em linearly unstable},
so that it does not rely on other mechanisms to reach the high
amplitudes necessary to explain the LFQPO.\\
The global structure of these modes is
formed by spiral waves travelling back and forth between the
inner disk edge (or its center) and their corotation
radius\footnote{not to be confused with the radius of corotation with
the central object}, where their angular phase velocity is equal to
that of the gas or stars. The mode
frequency is typically (depending on the profiles of various
quantities) .1 to .3 times the rotation frequency of the gas at the
inner disk edge, placing the corotation radius at $\sim$ 2.5 to 4
times the inner radius. It is most unstable when the magnetic field is near
equipartition, \ie when the plasma $\beta=8\pi p/B^2$ (the ratio of
thermal to magnetic pressure) is of the order of 1, and requires that
the radial gradient of the magnetic field be sufficient:\\
\begin{equation}
\frac{d}{d\ln r}\ln\left(\frac{\kappa^2}{2\Omega}\  
\frac{\Sigma}{B^2}\right) > 0
\label{eq:crit}
\end{equation}
where $\kappa$ and $\Omega$ are the epicyclic and rotation frequencies
(with $\kappa=\Omega$ in a keplerian disk), $\Sigma$ is the surface
density of the disk, and B the equilibrium magnetic field.  It is
noteworthy\footnote{J. Ferreira, private communication} that this
condition is always fulfilled in self-similar MHD models of jets
(\citet{BP82}, \citet{Lov86}, \citet{PP92}). \\
The restricted radial extension, 
and the condition that $\beta\sim 1$,
make the AEI a complement to the Magneto-Rotational Instability (MRI)
of \citet{BH91}. The MRI is a {\em local} instability (although modes
are also possible, see \citet{CP98}), and thus can occur throughout a
sufficiently ionized disk. On the other hand it is restricted to
weakly magnetized disks, \ie with $\beta>1$.\\
Figure \ref{fig:AEI} shows a schematic
description of the mode structure in and above the disk. The spirals are density waves, \ie sound waves whose propagation is
modified by differential rotation, by the Coriolis force (resulting in
epicyclic motion) and, if they apply, by self-gravity and the Lorentz
force. They become unstable by extracting energy and angular momentum
from the inner region of the disk (thus causing accretion) and finding
a way to transfer them outward: either (this is known as the swing
mechanism) to another spiral wave, traveling outward beyond
corotation; or (by the corotation resonance) to a Rossby wave at the
corotation radius: this is the main amplification mechanism for the AEI.\\
The AEI owes the `E' in its name to a unique characteristic:
as mentioned above, it grows by extracting energy and momentum from
the disk and transferring them to a Rossby vortex. In a purely
hydrodynamical or self-gravitating disk the process stops there, and
this led \citet{NGG87} to speculate, in the case of the PPI, that the
corotation resonance might saturate non-linearly when no more angular
momentum can be absorbed at corotation. However the AEI applies to
disks threaded by a poloidal field. Thus the footpoints of the
magnetic field lines anchored in the disk are submitted to a twisting
motion by the vortex, and this twist can now be propagated along the
field lines as Alfv\'en waves. It was suggested by \citet{TP99}, and
proven by \citet{vt02}, that these Alfv\'en waves could carry to the
corona of the disk a substantial fraction of the energy and momentum
extracted by the instability. This contrasts with all other disk
instability mechanisms, or with the turbulent viscosity model of
\citet{SS73}, where the energy and momentum are transported {\em
radially} outward and would need an additional mechanism to be
redirected upward. This gives the AEI the potential to directly feed
the winds or jets observed in all types of accreting
sources.\\
Note however that, at the present stage of theoretical work, we have
only shown that the energy and momentum are emitted upward as Alfv\'en
waves. There remains to show how these waves can deposit their energy in the
corona, to really produce a jet. Work in progress already
allows us to show how the instability can {\em heat} the corona, thus
permitting the Inverse Compton effect for
the high-energy tail of the source spectrum.\\
 Theory shows \citep{TPC92} that, as for galactic spirals, the
 structure of the AEI is essentially constant across the disk
 thickness. \citet{CT01} have used this property to perform numerical
 simulations, similar to early simulations of disk galaxies,
 considering an infinitely thin disk in vacuum. The drawback is that
 these simulations can describe neither the MRI (which develops within
 the disk thickness) nor the coupling of the AEI with Alfv\'en waves
 in the corona. But their relative simplicity allows very long
 simulations (tens of rotation periods), necessary to show the full
 non-linear development of the instability. These simulations are in very good agreement with the theoretical predictions.\\
\subsection{AEI and QPO}\label{subsec:QPO}
Our identification of the AEI as a candidate to explain the QPO starts
from its frequency, and relies on various other elements:
\begin{itemize}
\item The frequency of the AEI for the 1-armed mode (usually the most unstable) is typically .1~to .3 times the keplerian rotation
frequency at the inner edge of the disk, \Oint. This is consistent with the observed frequency of the LFQPO (typically in the range 1-10 Hz in 
black-hole binaries) and the inner radius given by spectral fits. They are observed to be correlated in \GG{} \citep{MMR99}. Furthermore \citet{Psa99} have found that, across a large number of sources of various types, the QPO frequency is correlated with a higher-frequency QPO believed to be of the order of \Oint.
\item The AEI is an instability, \ie it grows spontaneously and does not need
an external excitation, or the formation and long-term survival of
blobs of gas, to reach a large amplitude. Furthermore, and by analogy
with galactic spirals, we expect it to form a long-lived,
near-steady-state feature, described in the galactic context as a
Quasi-Stationary Spiral Structure - and indeed this is what we observe in our
simulations. This gives it a clear potential to yield the
observed Quasi-Periodic Oscillations. 
\item The LFQPO is so frequent in the `low-hard' state of black-hole 
binaries that it has been dubbed the `ubiquitous QPO'
\citep{Mar99}. For a detailed overview of the QPO properties in \GG,
see \citet{MMR99}, \citet{Rei00}, \citet{Tom01}. In the most common spectral 
analysis of the X-ray emission of these sources, as a multitemperature black body from 
the disk and a power-law tail from inverse Compton emission in the corona,
this state is characterized by a weak or inexistent disk emission, and
a dominant coronal one. This is to be expected from the AEI, since one
of its characteristics is that the accretion energy is transported
away by waves rather than dissipated locally to heat the disk, as
assumed in the standard model of turbulent viscosity
\citep{SS73}. Thus at comparable accretion rates the AEI should give a
lower disk emission and a more active corona than other instability
mechanisms.
In the course of the work discussed below, where we considered
relativistic effects, we found possible evidence for this.  We showed
that, at the transition between the high and the low state, during a
$\sim$ 30 minutes cycle of \GG, the QPO appeared just {\em before} the
spectral transition \citep{RVT02}: as shown in figure 7 of that paper, while the count rate decreases smoothly, the transition is marked by a sudden drop of the color temperature, which in one case is seen clearly to occur around 30 seconds after the apparition of the QPO.
Although it is presently limited to one
observation, this means that, if there is a causal relation between
the QPO and the transition to the low state, it is opposite to the one
usually assumed: it is {\em the apparition of the QPO} which causes
the transition, by stopping the heating of the disk and sending energy
to its corona.
%
%
\item In \citet{VRT02} and  \citet{RVT02} we have given a possible explanation for the observation, by \citet{Sob00}, that in the
microquasar \GRO{} the correlation between the color radius (taken as an indication of the disk inner radius $r_{int}$) and the LFQPO
frequency was opposite to the one usually found. Our interpretation is based on the sensitivity of the AEI  to relativistic effects on the rotation curve, when the inner disk edge approaches the Last Stable Orbit. It can of course be taken only as tentative, given the uncertainties in fitting the observed emission with a standard model of a multitemperature black body plus a power law. It is however remarkable, given these uncertainties, that our best fits indicate black-hole mass in agreement with independent measurements.
\item As a by-product of  our analysis of \GRO{} we also discussed how, when spectral 
fits give an anomalously low value for \rint {}
(sometimes smaller than the Schwartszchild radius of the black hole),
and if one retains the analysis in terms of a multicolor black-body contribution to the spectrum, 
this indicates emission from a small area at high
temperature. It might thus correspond to a hot spot, at the spiral
shock resulting from the AEI in the disk. \citet{Rod02b} found
possible evidence for this in \GG, by a determination of the energy
spectrum of the QPO at high energies.
\end{itemize}
\section{The variability of \GG}\label{sec:GG}
\begin{figure*}
\epsscale{2}
\centering
\plotone{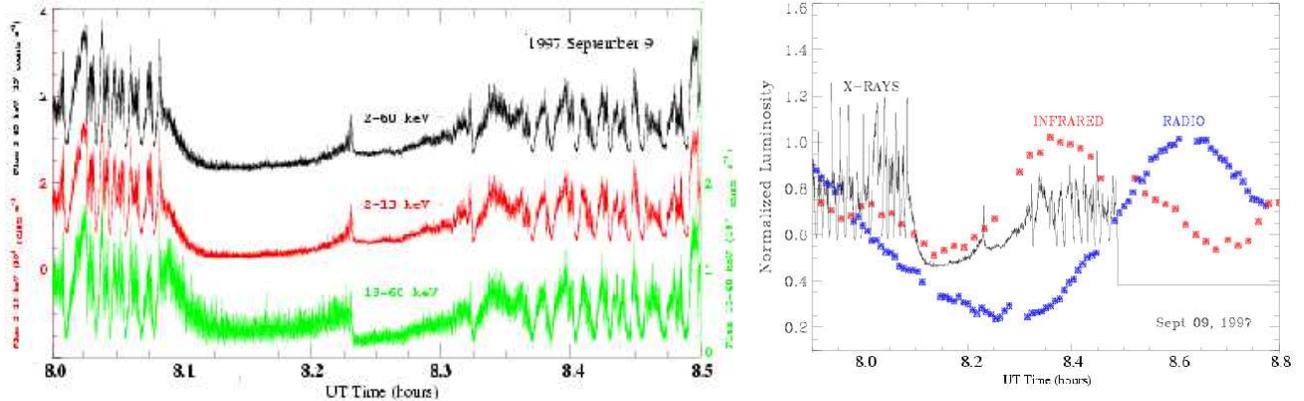}
\caption{A cycle ($\beta$ state of \citet{Bel00}) of \GG, observed on 
Sept. 9, 1997 \citep{Mir98, Cha98}. Top: the X-ray emission in various
bands. Bottom: total X-ray (black), IR (red) and radio (green)
emissions.
\label{fig:GG}}
\end{figure*}
\cite{Bel00} have shown that the variability of \GG{} could be reduced to 12 basic 
classes (noted by greek letters), which again could be
analyzed as successions of three basic states, {\em A, B} and {\em
C\/}. They also found that, after a steady hard class (noted $\chi$),
the most frequent one was class $\beta$, marked by the $\sim$ 30 minutes
cycles
we are interested in. These cycles, a
typical example of which is shown in figure \ref{fig:GG} \citep{Mir98,
Cha98}, have been the object of multi-wavelength observations which
contain a large number of informations. X~rays show first a high and
soft state (until time 8.1 in figure \ref{fig:GG}), then a rapid
transition to a low and hard state; at time 8.23 an intermediate spike
occurs, also marked by a sudden drop in the high energy (`coronal')
emission. The low-energy (`disk') emission keeps recovering until time
8.33, where the source transits back to the high state.  IR and radio
observations show a strong emission, after some delay which is
consistent with synchrotron emission from an adiabatically expanding
blob ejected at or near the intermediate spike. \citet{Eik98} analyze
more such cycles, confirming this association between the spike and
the blob ejection. More details can be found in \citet{Fen97},
\citet{Mir98}, \citet{Fen98}, \citet{Dha01}.\\
\citet{Mar99} performed a detailed timing analysis of the same X-ray 
data. They found that the low-frequency QPO was prominent, from the
transition to the low state until the spike. From spectral fits they
found that the inner disk radius \rint{} was near a minimal value at
the transition to the low state; it then increased progressively to a
maximum and decreased again until, just before the spike, it was back
to its minimal value. During the same period of time the QPO
frequency started from a maximal value, decreased and then grew again,
with variations opposite those of \rint. Since this particular
observation had been intensively analyzed and discussed, it is the one
we chose to analyze in our more recent work \citep{RVT02}. \\
The longer-term variability of \GG{} is more of a puzzle, as it
exhibits the 12 classes described by \citet{Bel00}. We will return to
this in section \ref{subsec:longterm}.
\section{Magnetic floods}\label{sec:flood}
\subsection{The transition to the low state}\label{subsec:transition}
Our Magnetic Flood scenario starts from the assumption, following the
discussion in section \ref{subsec:QPO}, that the AEI is indeed at the
origin of the QPO; we then wonder what this may tell us about the
physics of the inner region of the disk, and seek in the properties of
the AEI an explanation for the observations discussed in section
\ref{sec:GG}. As already mentioned, the low state associated with the
QPO can be understood in this context since the AEI transports the
accretion energy by spiral waves, rather than depositing it locally to
heat the disk. \\
Let us now consider what can explain the apparition of the QPO. It
should mean that an instability threshold is crossed: either the
radial profiles have evolved so that condition (\ref{eq:crit}) becomes
true, or the plasma $\beta$ becomes of the order of 1. We prefer
the second hypothesis for two reasons:
\begin{enumerate}[(i)]
\item The power density spectrum (PDS) of the source changes radically at the 
transition: it is close to a power-law in the high state (see \eg \citet{Rei03}), while in the low
state it shows broad-band noise at low frequencies, with a break just below the
QPO frequency. In the high state, we assume that turbulence in the
disk is due to the Magneto-Rotational Instability (MRI) of
\citet{BH91}, which is not sensitive to the radial profiles but
precisely has the property to be suppressed when $\beta\le 1$. This
suppression would explain that, besides the apparition of the LFQPO,
the PDS (which we associate with turbulence in the disk)
changes so strongly at the transition.
\item Furthermore this would explain that the transition to the low
state is so sharp, since as soon as the AEI appears the disk cools
down, further reducing the gas pressure and thus $\beta$.
\end{enumerate}
A gradual reduction of $\beta$ during the high state thus becomes our
working hypothesis. Two reasons can explain it:
\begin{description}
\item[Advection:] When considering the magnetic field in a disk, one 
should distinguish between the `horizontal' and `vertical' (\ie
parallel and perpendicular to the disk plane) components of the
magnetic flux. The horizontal (azimuthal and radial) flux can easily change, \eg when a
horizontal flux tube is lifted by buoyancy and leaves the disk (this
is the elementary mechanism of the Parker instability), or when
horizontal flux is generated by a turbulent dynamo. \\
Lifting a vertical flux tube, on the other hand, will not change the
total flux threading the disk. The only way to do this is, either by
reconnection with the magnetic structure in the inner hole surrounding
the central object (this will be discussed in section
\ref{subsec:centralmag}); or by letting the vertical flux diffuse
radially outward against the accretion flow, which may not be an easy
process, with advection competing against the turbulent magnetic
diffusivity.
\item[Dynamo:] The second possibility is that of a turbulent dynamo, 
associated with the MRI, which would create vertical flux of opposite
polarities in the inner and outer disk regions.
\end{description}
These possibilities ultimately depend on the transport properties of
the MRI; at this stage, numerical simulations have not been able to
give clear conclusions about the radial transport of vertical magnetic
flux by the MRI, \ie a turbulent magnetic diffusivity together with a
turbulent viscosity. Dynamo action has been observed by
\citet{Brand95} but not yet fully characterized. On the other hand,
the MRI is linearly an ideal MHD instability, and as such should cause
accretion of the magnetic flux {\em together} with the gas. Magnetic
diffusivity is involved in the non-linear evolution of the MRI, and
should break this property. To get an answer on this, we may have to
wait for more complex simulations, permitting in particular the
magnetic flux to cross freely the boundaries of the computational
domain. Or we could find an indication in the observation of strong
magnetic fields in the inner few tens of parsecs of the Milky Way:
there the field is vertical and measured in milligauss, rather than
horizontal and measured in microgauss elsewhere in the
disk. \citet{Cha00} have argued that this may be fossil magnetic flux,
advected with the gas during the history of the Galaxy. \\
Thus further studies are needed to assess the fraction of its embedded
magnetic flux the gas is able to transport inward. Whatever the
result, that fraction will accumulate near the central object and we
consider it as very likely that this will cause $\beta$ to gradually
decrease in the inner region of the disk. A final argument for this is
that MHD models of jets (\citet{BP82}, \citet{Lov86}, see
\citet{CF00}{ } and references therein) all require an `hourglass'
magnetic geometry, \ie vertical flux threading the disk, and
$\beta\sim 1$. These are axisymmetric equilibrium models, which as
mentioned previously are unstable to the AEI. Our work can thus be
considered as an elaboration on this line of models, assuming that
turbulence in the disk self-consistently generates this
configuration.\\
We also note that the accretion of gas can noticeably change the mass
of the central object only on a very long term. The same applies to
the magnetic flux of a neutron star. On the other hand the magnetic
flux threading a black hole is not anchored in the black hole, but is
due to currents near the inner edge of the disk. A secular evolution
of this flux may thus have a much stronger effect on the disk
behavior.
\subsection{The low/hard state}\label{subsec:low}
Following our scenario, in the low state the AEI transports the accretion energy outward to
its corotation radius; a sizable fraction of it can then be
transported to the corona, as shown by \cite{vt02}, so that the disk
cools down and the corona becomes more active. The rest of the energy
is dissipated in the disk at a spiral shock, forming a hot spot as
shown by the numerical simulations of \citet{CT01}; indications for
such a hot spot, associated with the QPO in \GG, have been discussed by
\citet{Rod02b}. \\
From the transition to the low state, the inner radius of the disk is
observed to move gradually outward; this may be the result of the
equilibrium between the standard, optically thick disk and whatever
lies between it and the black hole: a radiatively
inefficient disk, \eg an ADAF (in which case the inner radius of the
optically thick disk we are concerned with is set by the conditions of
transition to the ADAF), or \eg the force-free magnetic structure
associated with the \citet{BZ77} process, formed by a current ring at
the inner edge of the disk. In that case the change in \rint{} would
be set by the MHD equilibrium between this structure and the disk,
whose pressure has changed at the transition.\\
Then after some time \rint{} is observed to decrease again, until it
returns to the minimal value observed in the high state. This may be
attributed to the viscous refilling of the disk, as indicated by the
correlation \citep{Bel97} between the maximal value reached by \rint{}
(during this or other types of events) and the time it takes to return
to its minimal value. During this evolution the QPO frequency first
decreases, then increases, leaving its ratio to the keplerian
frequency at \rint{} in the range predicted by the AEI.

The low-hard state stops when \rint{} has returned to near its minimal
value, which it is of course tempting to identify with the Last Stable
Orbit (LSO). This is when the intermediate spike occurs, presumably
causing the relativistic ejections. \citet{Eik03} show that this is
consistent with a reconnection event. We will not discuss here their
`Magnetic Bomb' model, which involves a specific configuration for the
central magnetic structure connecting the black hole with the inner
edge of the disk. Whatever this configuration is, reconnection should
cause the ejection of a large part of the corona, as seen in figure
\ref{fig:GG} where after the spike the hard X-ray flux has
significantly dropped, which may indicate the ejection of part of the
corona \citep{Cha98}.\\
Reconnection destroys magnetic flux: this will allow the disk to
return to a state of low magnetization, and thus to a high
state. However this does not occur right away, and we should also seek
an explanation for this delay, which extends form the spike to time
$\simeq 8.32$ in figure \ref{fig:GG}.
\subsection{Belloni's states}\label{subsec:states}
\begin{figure*}
\centering
\epsscale{1}
\plotone{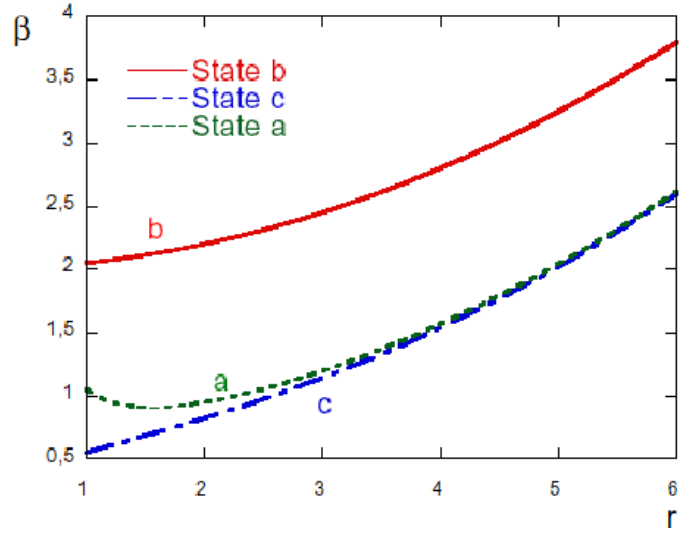}
\caption{The radial profile of $\beta=8\pi p/B^2$ in Belloni's three basic states, 
as predicted in our scenario. In state {\em B} the source is weakly magnetized, and in a classical
`high-soft' state. In state {\em C} it is more strongly magnetized and subject to the AEI, producing the LFQPO and leading to a `low-hard' state. In state {\em A} reconnection has weakened the magnetic field in the inner disk region,
keeping $\beta < 1$ but with a profile immune to the AEI.}
\label{fig:Beta}
\end{figure*}
\begin{figure*}
\epsscale{.75}
\plotone{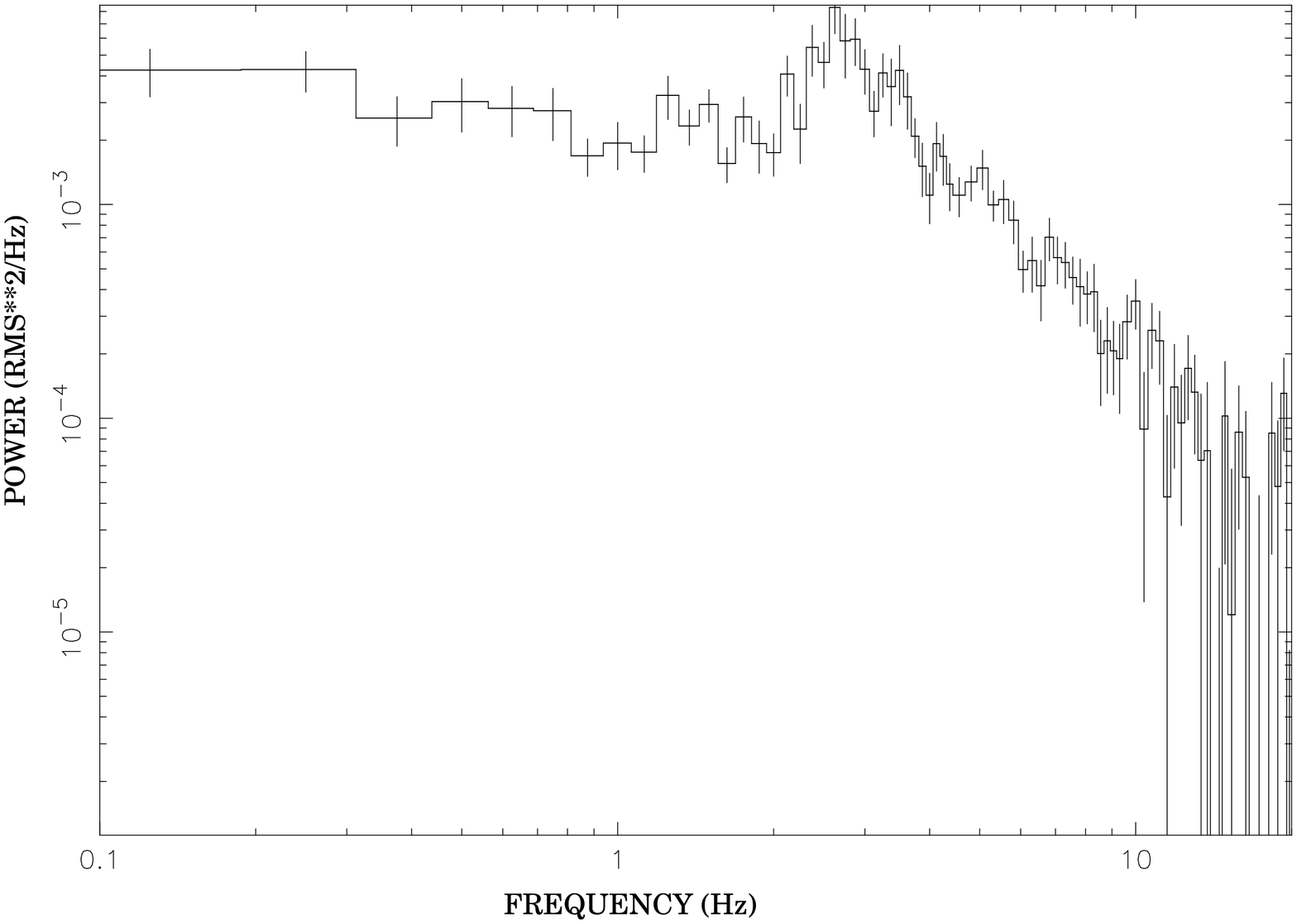}
\plotone{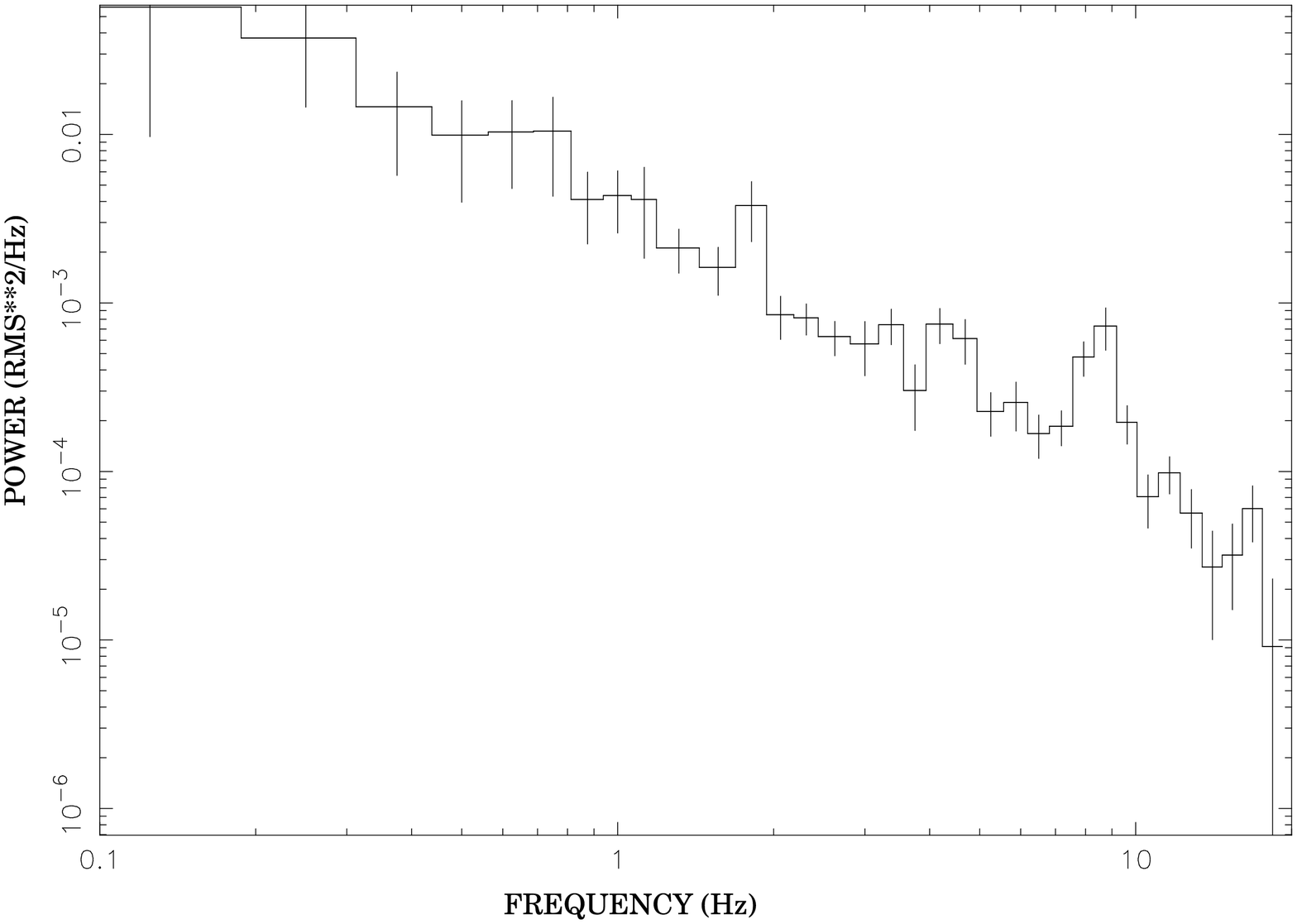}
\caption{Power Density Spectra taken while the source is in state {\em
C}. Upper panel: PDS taken at the bottom of the long dip (about 236~s
duration) of figure \ref{fig:GG}, while the source is already in the
{\em C} (hard) state. Bottom panel: PDS taken during an entire short
dip (53~s duration) where the source briefly transits to state {\em C\/}
while it is generally in the high oscillation phase.
 \label{fig:dips}}
\end{figure*}

The answer might lie in the work of \citet{Bel00}, who have shown that
the twelve classes of variability of \GG{} they had identified could
be reduced to alternances between three basic states, labelled {\em A,
B\/} and {\em C}, identified by the position of the source in a
color-color diagram. This allows a much more detailed view of the evolution of the source. 
State {\em B\/} is a conventional high-soft state,
with strong disk emission and weak coronal one; state {\em C\/}  is a
low-hard state, with weak disk and relatively strong corona (high
hardness ratio), while state {\em A\/}  is characterized by low emission
in all bands. The LFQPO is seen only in state {\em C}. The `high
state' during the cycles turns out, once analyzed on very short time
scales, to be actually a succession of these three basic states, while
the `low state' is {\em C\/} and after the spike the source is in state
{\em A}.\\
Following our line of inferences, we assume that state {\em C\/} is subject to the AEI: the inner region of the disk must thus be moderately magnetized with  $\beta\simlt 1$, and obey the instability condition (equation~\ref{eq:crit}). The cool disk and active corona, as well as the presence of the compact jet (see \eg \citet{Rei03}), are consistent with the presence of the AEI. \\
It is also natural to consider state {\em B\/} as a weakly magnetized ($\beta>1$) state subject to the MRI, which is our best candidate to produce turbulence leading to a local heating of the disk by the accretion energy (whether or not the effects of the MRI can be modeled by a standard $\alpha$ disk: see \eg \citet{BH98}).\\
We still have to understand in this context the nature of  state {\em A}. It cannot be a state where the inner region of the disk is weakly magnetized, since it would be subject to the MRI whatever the other disk parameters are,  and should thus show properties similar to state {\em B}.  It must thus have $\beta\leq 1$, and be stable to the AEI: this can occur if the
instability criterion of equation \ref{eq:crit} is not verified, \ie if the quantity $B^2/\Sigma$ does not decrease faster than $\Omega$ outward.\\
The association of the spike with a reconnection event makes this a
very likely possibility: reconnection, destroying magnetic flux at the
inner disk edge, should weaken or even reverse the radial magnetic
field gradient in the inner disk region, while leaving the disk cool
enough (in the absence of turbulence) to keep $\beta\leq 1$. It could
thus make both the AEI and MRI stable, leading to state {\em
A}. Accretion proceeding from the outer part of the disk, and
gradually refilling the inner disk with weakly magnetized gas, would
after some time allow the observed return to state {\em B}. Figure
\ref{fig:Beta} sketches the corresponding profiles of $\beta$ in
Belloni's three basic states, in the inner disk region.\\
This immediately leads to a possible explanation for the striking observation of
\citet{Bel00}, that all transitions between their basic states are seen
{\em except} the transition from state {\em C\/} to {\em B\/}: in our scenario, state {\em C\/} means that the disk has accumulated a critical amount of vertical magnetic flux in its inner region. Once it is in that state, it can decrease its
magnetization {\em only} by a reconnection event, leading to state
{\em A}. Then accretion from the outer regions replenishes the inner disk with low magnetization material, allowing the MRI-unstable region to progress inward until it reaches the inner disk edge, seen as a return to state {\em B}. This would then explain the smooth transition observed from state {\em A\/} to {\em B}, in contrast with the sharp transitions from {\em C\/} to {\em A\/} (due to reconnection) and from {\em B} to {\em C\/}: there,  as discussed in section \ref{subsec:low}, the transition is sharp because as $\beta$ falls below 1, the MRI stops, so that the disk cools down, further decreasing $\beta$.\\
This has then suggested us to seek an interpretation for the LFQPO sometimes seen during short dips
occurring while the source is in the `high/soft' state. This is discussed and shown in particular in figure 1 of \citet{Mar99}: they find that the high state of a class $\beta$ cycle is interrupted by such dips, where the source is in a spectral state similar to the long dips (this, shown by black intervals in the center bar of their figure,  must thus be  state {\em C\/}), and the LFQPO appears. Following our scenario, this occurs at times where the
radial profile of $B$ still leaves $\beta>1$, while magnetic flux is
slowly accreted to the inner disk region. A possible explanation,
following our scenario, would be that at these times $\beta$ in the
inner disk is not much larger than one, so that the AEI can appear but
that these episodes remain transient until a necessary condition is
met to let the long dips occur. This condition would then certainly
concern the global magnetic field profile, allowing during the low
state the outward motion of the inner disk edge which characterizes
the long dips.\\
Seeking an indication in favor of this explanation, we have examined
the Power Density Spectrum of the source, comparing a short dip and a long
one. The results are shown in figure \ref{fig:dips}. The first
spectrum, obtained near the minimum of the long dip of figure
\ref{fig:GG}, shows the characteristic aspect with band-limited noise
at low frequencies, a break, and the QPO (here near 3 Hz). The second
spectrum is taken during the short dip shown at time $\sim 2450-2500$ in figure 1 of \citet{Mar99}, and is indeed very
different: it shows a power-law spectrum, as usually found in the
low-high state (state {\em B}), to which the QPO is superimposed near
9 Hz. We would expect the break at a frequency higher than 1 Hz, while the interval analyzed is 53 seconds long so that, although the disk properties are more transient here than during the long dip, we believe that this is not due to the limited time window. On the other hand interpretation is difficult, since the disk emission is more
present here than during the long dip (as the disk is hotter). However, and although this would need to be analyzed in more details over a number of occurrences, 
it might thus indicate that these occurrences of state {\em C} are
in fact slightly different; in our interpretation, during the short
dip the formation of the QPO does not manage to quench the MRI
turbulence associated with the power-law PDS,
characteristic of the high/soft state.
\subsection{The central magnetic structure}\label{subsec:centralmag}
Reconnection at the inner edge of the disk is the best explanation for
the spike leading to the relativistic ejections; it would merge
magnetic field lines from the disk and field lines from the central
magnetic structure surrounding the black hole. We consider here the
same magnetic structure involved in the Blandford-Znajek process
\citep{BZ77}, whether or not this process is efficient enough to
generate a jet, in its force-free initial form or in more elaborate
ones such as used by \citet{Eik03}. Their model of the global magnetic
configuration, involving opposite current loops in a torus around the
black hole, is not unique and different from ours where the disk is
threaded by open field lines. One might probably, rather than
reconnection, find alternative explanations based on other types of
large-scale magnetic events, but reconnection is clearly the first
possibility to consider. However for this to occur the poloidal
magnetic flux accreted with the gas and the one trapped in the central
magnetic structure, in the inner hole surrounding the central object,
must have opposite signs.  \\
This would mean that the classes of variability where ejections occur
can happen only when the poloidal fluxes in the disk and this
structure are antiparallel, and thus that during certain periods the
source is in that configuration, while during other periods the fluxes
have the same sign. This would lead to a dichotomy, similar to the one
observed at the interface between the terrestrial magnetic field and
the one carried by the solar wind: it is only when these fields are
antiparallel that reconnection is observed, while in the parallel
configuration more complex behaviors occur.  \\
Thus, although it may seem to be a far-fetched assumption for the
interface between the disk and the central magnetic structure, it is
hard to escape if reconnection is involved. One also has to remember
that the central magnetic structure is not anchored in the black hole,
but in a current ring at the inner edge of the disk; the poloidal flux
it contains is thus not a constant, as in neutron stars whose flux can
vary only over very long periods. Here the trapped flux is the sum of
all the magnetic flux accreted during the whole history of the source,
while the gas itself will end up in the black hole. \\
On the other hand, the flux advected with the gas may also change over
time: as discussed in section \ref{subsec:transition}, this flux may
have its origin either in the companion star the gas comes from, or in
a turbulent dynamo in the disk. In both cases the sign of this flux is
expected to vary, although the characteristic time scale of this
variation is difficult to assess (there is for instance still no
quantitative explanation for the duration of the 22-year cycle of the
Sun, associated with such field reversals, and convection must be much more complex in a companion star stirred by tidal effects). We will discuss below how the time scale
observed in \GG, which we would associate with this long-term
evolution, is of the order of one to a few years. This is certainly
not unreasonable for either process. \\
Thus the processing of magnetic flux in the disk and the central
magnetic structure would be responsible for two time scales: a short
one, responsible for the $\sim$ 30 minutes cycles, corresponding to the
accumulation and release of magnetic flux in the inner region of the
disk; and a much longer one, corresponding to changes in the sign of
the flux advected with the disk, and due to changes at the source of
this flux, be it a dynamo in the disk itself or the companion star.
\subsection{Long-term behavior}\label{subsec:longterm}
\begin{figure*}
\centering
\plotone{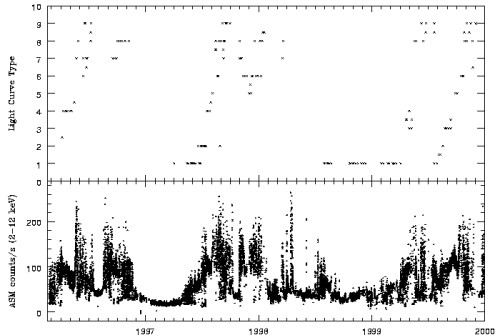}
\caption{This figure, reproduced from \citet{Fitz00}, shows: 
lower plot, the X-ray light curve of \GG{} during the period studied;
upper plot, the history of the source in terms of their classes of
variability. Each class corresponds to a type of variability of the
source, in a manner similar to \citet{Bel00}. Clearly they occur in a
well-defined and repetitive order.  \label{fig:FitzModes}}
\end{figure*}
%
That there is an unknown parameter, besides the accretion rate,
regulating the long-term behavior of \GG{} has long been
suspected; from a statistical analysis \citet{Gre03} have found a 12-17 days time scale which might ba associated with the orbital motion of the binary. More recently \citet{Rau03} found in its hard X-ray and radio fluxes a 590 days period. Their preferred explanation involves the precession of a warp in the disk, but it they find it difficult in this context to understand that this periodicity is seen in the coronal emission only and not in the soft X-ray flux from the disk. We will first note here that a turbulent dynamo model would appear as a better candidate since the magnetic field in the disk directly affects dissipation in the corona. As mentioned in section \ref{subsec:centralmag},  such a period would not be unrealistic for field reversals in the disk or even in the companion star.\\
However we also wish to mention unpublished work by 
\citet{Fitz00}, which also indicates order in the long-time variability of \GG{}, including its disk emission: examining over 350 RXTE observations of \GG{} between 1996 and 1999, they first defined nine basic classes of variability
of \GG, analogous (and generally similar) to the twelve ones independently found by \citet{Bel00};
they ordered these classs (numbered 1 to 9) by various criteria, the
main one being the direct observation of transitions between two
classes. Then they plotted, over this period
of nearly four years, the observed class of variability. Their result,
reproduced in figure \ref{fig:FitzModes}, shows that the source
follows a regular pattern, going from class 1 to class 9 before it
returns to a steady state (not included among the nine classes, and not
shown in figure \ref{fig:FitzModes}). Then after some time it starts
over at class 1. They also find that there is no strong correlation
between this and the total flux of the source. We cannot reproduce
here their analysis, which would deserve to be repeated in more details and in terms of the 12 classes of \citet{Bel00}, since they have become the standard description for the variability of \GG; a comparison between this and the periodicity of \citet{Rau03}{} would also of course be necessary. We will only note that these classes are very similar to Belloni's, and that the $\sim$ 30 minutes cycle (class $\beta$) is their class~8. Our main interest here is the quite unexpected existence of such a regularity in the long-term behaviour of \GG, which must be considered as an outstanding challenge to any model of accretion in this source. \\
 In section \ref{subsec:centralmag} we concluded that, if reconnection
 events are involved in the spike of the $\sim 30$ minutes cycles and in the relativistic
 ejections, this is most likely to occur during periods where the
 poloidal fluxes in the disk and the central hole have opposite signs;
 we must then expect the source to be in this configuration
 (antiparallel vertical fields) part of the time, and in the opposite
 one (with parallel fields) at other times. Changing from one
 configuration to the other requires a change in the sign of either
 the magnetic flux advected with the gas in the disk, or the total
 flux trapped in the central magnetic structure.\\
A change in the sign of the advected magnetic flux may result from a
field reversal in the dynamo generating the advected flux, in the
companion star or in the disk itself. Field reversals are to be
expected in turbulent dynamos, but not much is known about their
characteristic time scale. The orientation of the magnetic field 
should not affect the main body of the
disk (weakly magnetized anyway); on the other hand it should strongly
affect the interaction with the central magnetic structure, and thus
the accretion process in the inner region of the disk. On the long
term, a sufficient number of reconnection events might
eventually cancel the trapped flux, and change its sign. \\
Since we are elaborating on a model of the cycles, which is itself an
extrapolation of our candidate for the low-frequency QPO, we will not
try to go further in this direction. We however consider very possible
that the long-term evolution seen in figure \ref{fig:FitzModes} could
result from the processing of poloidal magnetic flux in the disk and
the central magnetic structure.
\section{Discussion}\label{sec:discu}
In this paper we have started by presenting our model for the
Low-Frequency Quasi-Periodic Oscillation of X-ray binaries,
particularly the microquasar \GG, based on the Accretion-Ejection
Instability predicted to occur in the inner region of magnetized
disks. Assuming that the AEI does expain the LFQPO, we have then
proceeded by extrapolation, seeking in the properties of the
instability an explanation for the observed behavior of the source.\\
This has allowed us to build up a scenario for the $\sim 30$ minutes cycles
(Belloni's class $\beta$) of \GG. We call this a `Magnetic Flood'
model, because it explains the cycles by the gradual accumulation of
magnetic flux in the inner region of the disk, and sudden release in
reconnection events producing the relativistic ejections. The scenario
is consistent with both the observations and the physical properties
of the instability, \ie its instability conditions, and its transport
of accretion energy outward and to the corona (so that the disk cools
down and the corona is energized when the AEI is present). This allows
us to find a natural explanation for the observation that the
apparition of the QPO occurs {\em just before} the transition to the
low state, so that the QPO may {\em cause} the transition, rather than
be its consequence.

Turning to the reduction, by \citet{Bel00}, of the variability of \GG{} to 12 
classes and then to 3 basic states, {\em A}, {\em B} and {\em C\/}, we then identify these states of the inner disk region as
\begin{description}
\item{\bf state {\em B} :} a standard, weakly magnetized disk ($\beta > 1$) subject to the Magneto-Rotational Instability; the local deposition of accretion energy heats the disk resulting in  a high-soft state
\item{\bf state {\em C} :} a more (though moderately) 
magnetized state, where the poloidal field has become of the order of
equipartition with the gas pressure ($\beta\simlt 1$); in this state
the MRI is stabilized and the AEI unstable, producing the LFQPO,
reducing the disk heating and exciting the corona where a part of the
accretion energy is transported
\item{\bf state {\em A} :} a state which is still moderately
magnetized ($\beta \simlt 1$, preventing the MRI) but where magnetic
flux near the inner edge of the disk has been destroyed in the
reconnection; the resulting magnetic field profile does not allow the
AEI to be unstable, so that the inner disk is essentially calm until
accretion proceeding from larger radii increases $\beta$ and allows a
return to state {\em B}.
\end{description}
This allows us to give a possible explanation for the observation of
\citet{Bel00} that a direct transition from state {\em C} to state
{\em B} is never observed: magnetic flux has to be destroyed, leading
to state {\em A}, before accretion of gas from larger radii allows a
return to state {\em B}.\\
We have then explored how this could allow a further extrapolation, to
the other  classes of variability of 
\GG   {} and its behavior on longer time scales. For this we started
from the conclusion of our model, that the source must be part of the
time in a configuration where the vertical magnetic flux in the disk
and the one trapped in the central structure surrounding the black
hole are parallel, while at other times they are antiparallel. This is
in fact unescapable if the relativistic ejections are due to a large
scale reconnection event between the disk and the central magnetic
structure.  We confronted this with the unpublished work of
\citet{Fitz00}, which shows a surprising regularity in the 
manner \GG{} explores its different classes of variability. We are led
to suggest that this regularity in the long-term evolution is ruled by
reversals in the sign of the magnetic flux advected to the inner disk
region. This flux may result from a dynamo in the companion star or in
the disk itself, and is eventually accreted to the central magnetic
structure, which may thus also reverse its sign periodically. \\
\acknowledgments 

The authors wish to acknowledge a large number of discussion, all
along the elaboration of this model, with T.Belloni, S. Chaty,
S. Corbel, Ph. Durouchoux, R. Fender, J. Ferreira, Y. Fuchs, R.N. Henriksen,
F. Mirabel, E. Morgan, M. Muno, and R. Remillard. They also wish to express special thanks to the anonymous referee, for a very constructive discussion leading to a sharp clarification of our arguments. JR acknowledges financial support from the French Space Agency (CNES).
 \clearpage \end{document}